\documentclass[10pt]{article}

\usepackage{amsmath}
\usepackage{amssymb}
\usepackage{float}
\usepackage{graphicx}
\usepackage{ltablex}

\usepackage{cite}

\usepackage{color} 
\usepackage{tabularx}
\usepackage{subfigure}

\usepackage[labelfont=bf,labelsep=period,justification=raggedright]{caption}

\makeatletter
\renewcommand{\@biblabel}[1]{\quad#1.}
\makeatother

\date{}

\begin{document}

\begin{flushleft}
{\Large
\textbf{Collective attention in the age of (mis)information}
}
\\
Delia Mocanu$^{1}$,
Luca Rossi$^{1}$,
Qian Zhang$^{1}$,
M\'arton Karsai$^{1,2}$
Walter Quattrociocchi$^{1,3}\ast$
\\
\bf{1} Laboratory for the Modeling of Biological and Socio-technical Systems, Northeastern University, Boston, MA 02115 USA
\\
\bf{2} Laboratoire de l'Informatique du Parall\'elisme, INRIA-UMR 5668, IXXI,  ENS de Lyon 69364 Lyon, France
\\
\bf{3} Laboratory of Computational Social Science, IMT Alti Studi Lucca, 55100 Lucca, Italy
\\$\ast$ Corresponding Author E-mail: walter.quattrociocchi@imtlucca.it
\end{flushleft}

\section*{Abstract}
In this work we study, on a sample of 2.3 million individuals, how Facebook users consumed different information at the edge of political discussion and news during the last Italian electoral competition. Pages are categorized, according to their topics and the communities of interests they pertain to, in a) alternative information sources (diffusing topics that are neglected by science and main stream media); b) online political activism; and c) main stream media.
We show that attention patterns are similar despite the different qualitative nature of the information, meaning that unsubstantiated claims (mainly conspiracy theories) reverberate for as long as other information.  Finally, we categorize users according to their interaction patterns among the different topics and measure how a sample of this social ecosystem (1279 users) responded to the injection of 2788 false information posts. Our analysis reveals that users which are prominently interacting with alternative information sources (i.e. more exposed to unsubstantiated claims) are more prone to interact with false claims.

\section*{Introduction}
The World Economic Forum, in its 2013 report \cite{Davos13}, has listed
the ``massive digital misinformation'' as one of the main risks for
the modern society.  People perceptions, knowledge, beliefs, and opinions
about the world and its evolution get (in)formed and modulated through
the information they can access, most of which coming from newspapers,
television \cite{Maxwell}, and, more recently, the Internet.  The world wide web, more specifically
social networks and micro-blogging platforms, have changed the way we
can pursue intellectual growth or shape ideas. In particular, large social
networks, with their user-provided content, have been facilitating the
study of how the economy of attention leads to specific patterns for the
emergence, production, and consumption of information
\cite{Lanhanm2007,Qazvinian2011,DAF13}. 

Despite the enthusiastic rhetoric about the ways in which new
technologies have burst the interest in debating political or social
relevant issues
\cite{Guillory2011,Bekkers2011,Moreno2011,Garcia2012,crespi,lippmann},
the role of the socio-technical system in enforcing informed debates and
their effects on the public opinion still remain unclear.  Indeed, the
emergence of knowledge from this process has been dubbed
\emph{collective intelligence}
\cite{Levy1999,Buckingham2012,Levy2000,Malone2007,Shadbolt2013},
although we have become increasingly aware of the presence of
unsubstantiated or untruthful rumors.  False information is
particularly pervasive on social media, fostering sometimes a sort
of \emph{collective credulity}.

In this respect, conspiracists tend to explain significant social or
political aspects as plots conceived by powerful individuals or
organizations \cite{Sunstein12}.  As these kind of arguments can sometimes
involve the rejection of science, alternative explanations are
invoked to replace the scientific evidence.  For instance, people who
reject the link between HIV and AIDS generally believe that AIDS was
created by the U.S.~Government to control the African American
population \cite{Bogart05,Kalichman09}. Since unsubstantiated claims are
proliferating over the Internet, what could happen if they were used as
the basis for policy making?

A multitude of mechanisms animate the flow and acceptance of false
rumors \cite{Kuklinski2000}, which in turn create false beliefs that are
rarely corrected once adopted by an individual
\cite{Garrett2013,Meade2002,koriat2000,Ayers98}.  The process of
acceptance of a claim (whether documented or not) may be altered by
normative social influence or by the coherence with the individual
system of beliefs \cite{Zhu2010,Loftus2011}. Nonetheless, several tools have
been recently designed to help users disambiguate misinformation and
false news \cite{Ennals:2010,mckelvey2013truthy}. On the other
hand, basic questions remain on how the quality of (mis)information
affects the economy of attention processes, concerning, for example, the
virality of information, its lifespan and the consumption patterns.

A large body of literature addresses the study of social dynamics on
socio-technical systems
\cite{Onnela2010,Ugander2012,Lewis2012,Mocanu2012,Adamic05thepolitical,kleinberg2013analysis};
here we consider the relationship between information sources and 
online political debates, limiting our investigation to
the period preceding the Italian elections of 2013, and focusing our
attention on the Italian Facebook groups formed around political
activism and alternative news sources.

We observed several interesting phenomena such as the proliferation of political pages and alternative
information sources with the aim to exploit the Internet peculiarities
to organize and convey the public discontent (with respect to the crisis
and the decisions of the national government).  Furthermore, we noticed
the emergence of very distinct groups, namely {\em trolls}, building
Facebook pages as a parodistic imitation of both alternative information
sources and online political activism. Their activities range from
controversial comments and posting satirical content mimicking
alternative news sources, to the fabrication of purely fictitious
statements, heavily unrealistic and sarcastic. Not rarely, these memes
became viral and were used as evidence in online debates from political
activists \cite{forconi_cirenga2013}.  Inspired by these lively and
controversial social dynamics, we addressed the quantitative analysis of
the interlink between information sources and political activism on the
web. In particular, we want to understand the selection criteria of
users mostly exposed to unsubstantiated claims.

This paper is structured as follows.  We will first introduce our methodology of categorizing the Facebook pages, by taking into account their self-description as well as the type of content they promote. We concentrate on alternative news sources, online political activism, and also on all the national main stream news journals that we could find to have an active page on Facebook. 
In the following sections, through thorough quantitative analysis, we show that the attention patterns when faced with various contents are similar despite the different qualitative nature of the information, meaning that unsubstantiated claims reverberate as long as other, more verified, information. Finally, we measure how the social ecosystem responded to the perturbation of false information injected by \textsl{trolls}. 
We find that a dominant fraction of the users interacting with the troll memes is the one composed of users preeminently interacting with alternative information sources -- and thus more exposed to unsubstantiated claims.
Surprisingly, consumers of alternative news, which are the users trying to avoid the main stream media 'mass-manipulation', are the most responsive to the injection of false claims.

\section*{Methods}

\subsection*{Case study and data collection}
The debate around relevant social issues spreads and persists over the web, leading to the emergence of unprecedented social phenomena such as the massive recruitment of people around common interests, ideas or political visions.
Disentangling the many factors behind the influence of information sources on social perception is far from trivial. Specific knowledge about the cultural and social context (even if online) in which they manifest is fundamental. 
Hence, inspired by the success of political movements over the Internet, we start our investigation focusing on the social dynamics around pages of political activism on the Italian Facebook during the 2013 electoral campaign.  On one hand, political activists conveyed the public discontent on the government and the economic conditions on a public arena; on the other hand, as the main stream media are considered to be manipulated, alternative information sources were free to disseminate news neglected by mainstream media or by science.
In addition, we notice the activity of an emerging group of users, namely {\em trolls}, producing caricatural versions of the stories diffused by alternative information sources and political activism pages. As an outcome of this period of observation, we compile a list of the most important and active Facebook pages of alternative information sources and political movements. 

The dataset is composed of 50 public pages for which we download all the posts (and their respective users interactions) in a time span of six months (from Sept 1st, 2012 to Feb 28th, 2013).
The entire data collection process is performed exclusively with the Facebook Graph API \cite{fb_graph_api}, which is publicly available and which can be used through one's personal Facebook user account.
The pages from which we download data are public Facebook entities (can be accessed by virtually anyone). Most of the user content contributing to such pages is also public unless the user's privacy settings specify otherwise. The exact breakdown of the data is presented in Table \ref{tab:data_dim}. We provide brief descriptions for each page in Supporting Information. 

The categorization of the pages is based on their different social functions together with the type of information they disseminate. The first class includes all pages (that we could verify) of main stream newspapers; the second category consists of alternative information sources - pages which disseminate controversial information, most often lacking supporting evidence and sometimes contradictory of the official news (e.g. conspiracy theories, link between vaccines and autism etc). 
The third category is that of self-organized online political movements -- with the role of gathering users to publicly convey discontent against the current political and socio-economic situation (i.e. one major political party in Italy has most of its activity online).

For all classes the focus of our analysis is on the interaction of users with the public posts -- i.e, likes, shares, and comments.

\begin{center}
\begin{table}[!h]
\centering
\begin{tabular}{l|c|c|c|c|c}
\bf {  }  & \bf {Total} & \bf {Mainstream News} & \bf {Alternative News} & \bf {Political Activism} \\ \hline 
Distinct users & $ 2,368,555 $  & $ 786,952 $ & $ 1,072,873 $ & $ 1,287,481 $ \\ 
Pages & $ 50 $ & $ 8 $ & $ 26 $ & $ 16 $\\
Posts & $ 193,255 $ & $ 51,500 $ & $ 92,566 $ & $ 49,189 $\\
Likes & $ 23,077,647 $ & $ 4,334,852 $ & $ 7,990,225 $ & $ 10,752,570 $\\
Comments & $ 4,395,363 $ & $ 1,719,409 $ & $ 935,527 $ & $ 1,740,427 $\\
Likes to Comments & $ 4,731,447 $ & $ 1,710,241 $ & $ 1,146,275 $ & $ 1,874,931 $\\
\end{tabular}
  \caption{ Breakdown of Facebook dataset. \textbf{Mainstream News}: all the national newspapers present on Facebook. \textbf{Alternative News}: pages which disseminate controversial information, most often lacking supporting evidence and sometimes contradictory of the official news.. 
\textbf{Political Activism}: gathering users to publicly convey discontent against the current political and socio-economic situation.} 
  \label{tab:data_dim}
\end{table}
\end{center}

Finally, we got access to 2788 post ids from a troll Page \cite{SemplicementeMe}. All of these posts are caricatural version of political activism and alternative news stories, with the peculiarity to include always false information. 
Despite the small dimension (7430 unique users, 18212 likes, 11337 comments and 9549 \textsl{likes to comment}) the page was able to trigger several viral phenomena, one of which reached $100K$ shares. We use troll memes to measure how the social ecosystem under investigation is responding to the injection of false information.

\section*{Results and Discussion}

\subsection*{Attention patterns}

\begin{figure}[H]
 \centering
       \includegraphics[width=.7\textwidth]{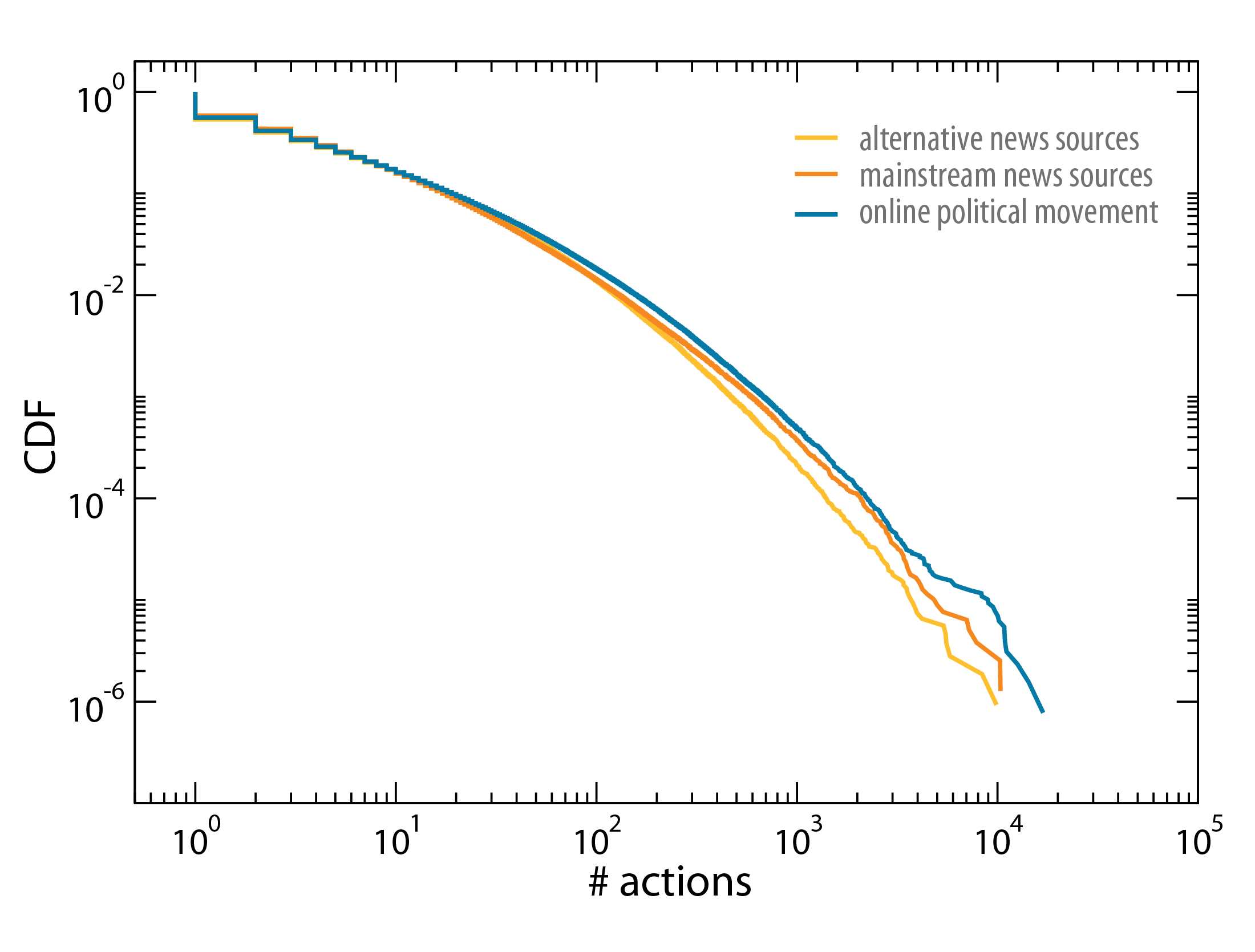} \label{fig:p_l}
\caption{\textbf{Users Activity}. Cumulative distribution function (CDF) of users' activity, grouped by page type. An action can be a like, comment, or \textsl{like to comment}. The distributions are nearly identical.}
\label{fig:activity_distribution}
\end{figure}

\begin{figure}[H]
 \centering
       \includegraphics[width=.7\textwidth]{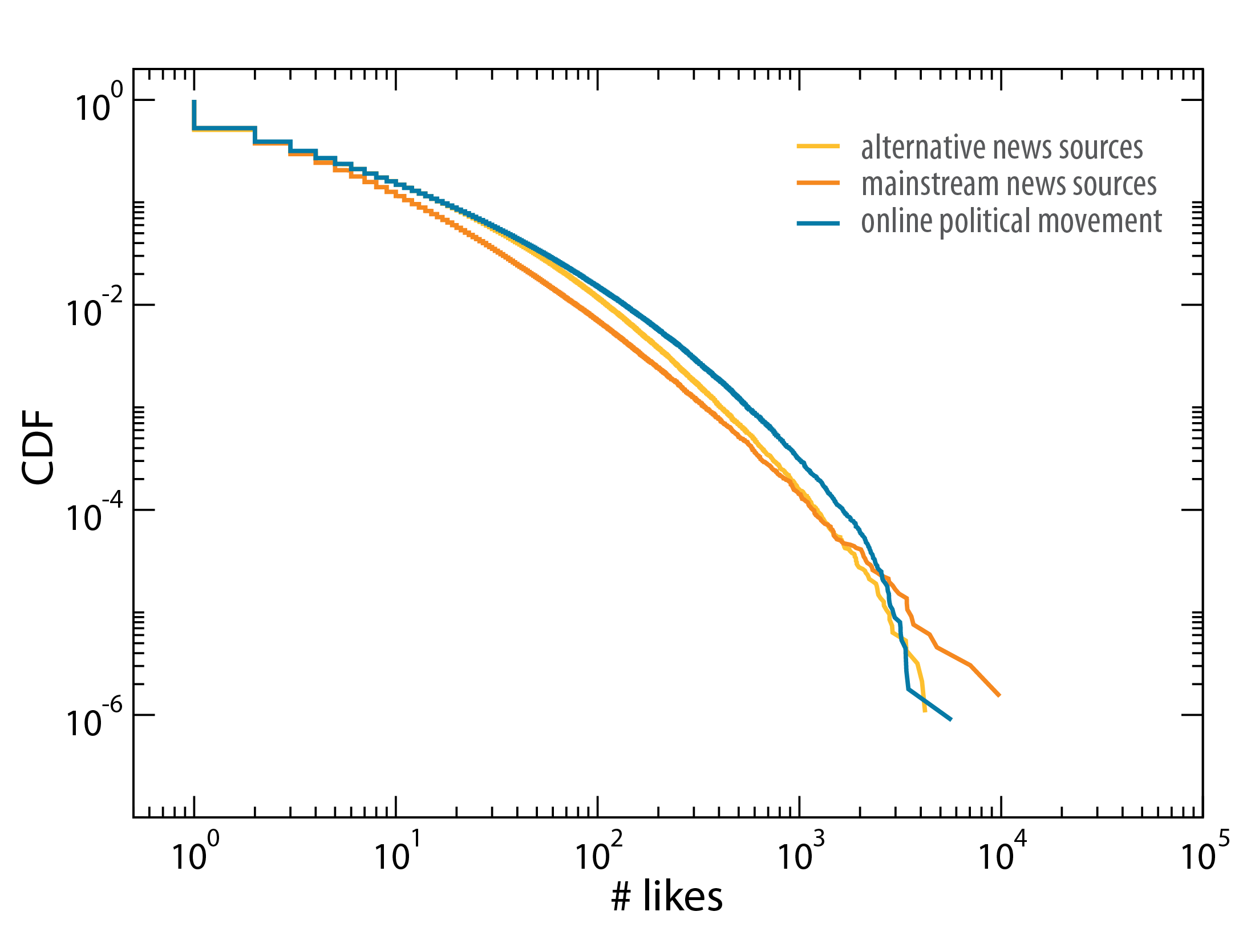} \label{fig:p_l}
\caption{Users' likes cumulative distribution function, grouped by page type.}
\label{fig:activity_distribution_likes}
\end{figure}

\begin{figure}[H]
 \centering
       \includegraphics[width=.7\textwidth]{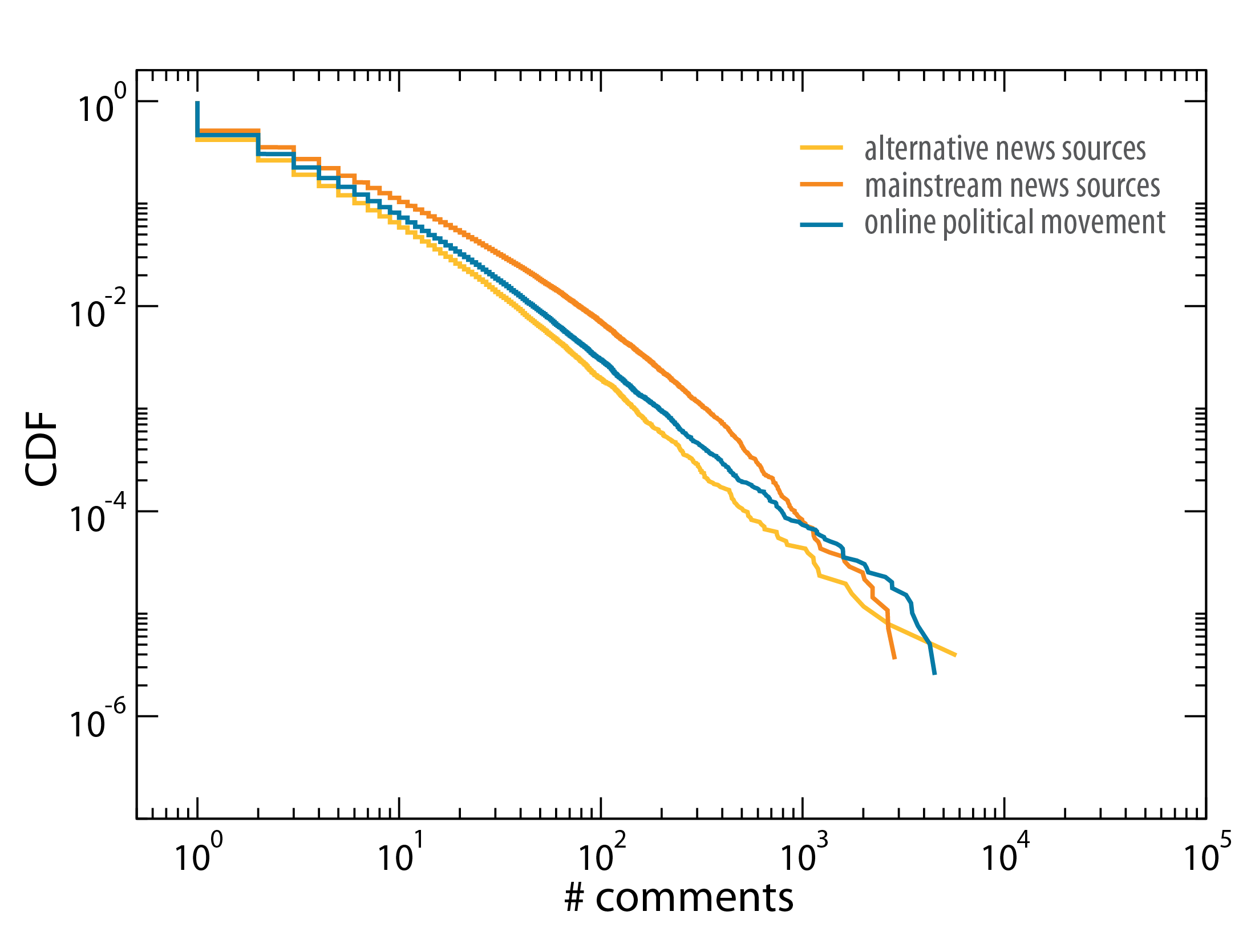} \label{fig:p_l}
\caption{Users' comments cumulative distribution function, grouped by page type.}
\label{fig:activity_distribution_comments}
\end{figure}

We start our analysis providing an outline of users' attention patterns with respect to different topics coming from distinct sources - i.e, alternative news, main stream media and political activism.  As a first measure, we count the number of interactions (comments, likes, or \textsl{likes to comments}) by users and plot the cumulative distribution function (CDF) of the users' activity on the various page categories in Figure \ref{fig:activity_distribution}.
CDF shows that user interactions with posts on all different types of pages does not present significant differences. The similarity is also conserved after further grouping comments and likes separately (see in Supporting Information Figure \ref{fig:activity_distribution_likes} and Figure \ref{fig:activity_distribution_comments}).

Here, the social response is not affected by the topic nor by the quality of the information. Posts containing unsubstantiated claims, or about political activism, as well as regular news, cannot be distinguished through simple statistic signatures based on user engagement patterns. These different topics reverberate at the same way in this ecosystem.

As the potential of memes to trigger discussions can be quantified through comments, in order to have a more precise picture of the users' attention, we zoom in to the level of posts. 
This level of resolution is useful to understand the temporal evolution of posts and for how long the debate on a topic persists, using the comments as a first-order approximation of the level of interest.  

\begin{figure}[H]
 \centering
       \includegraphics[width=.7\textwidth]{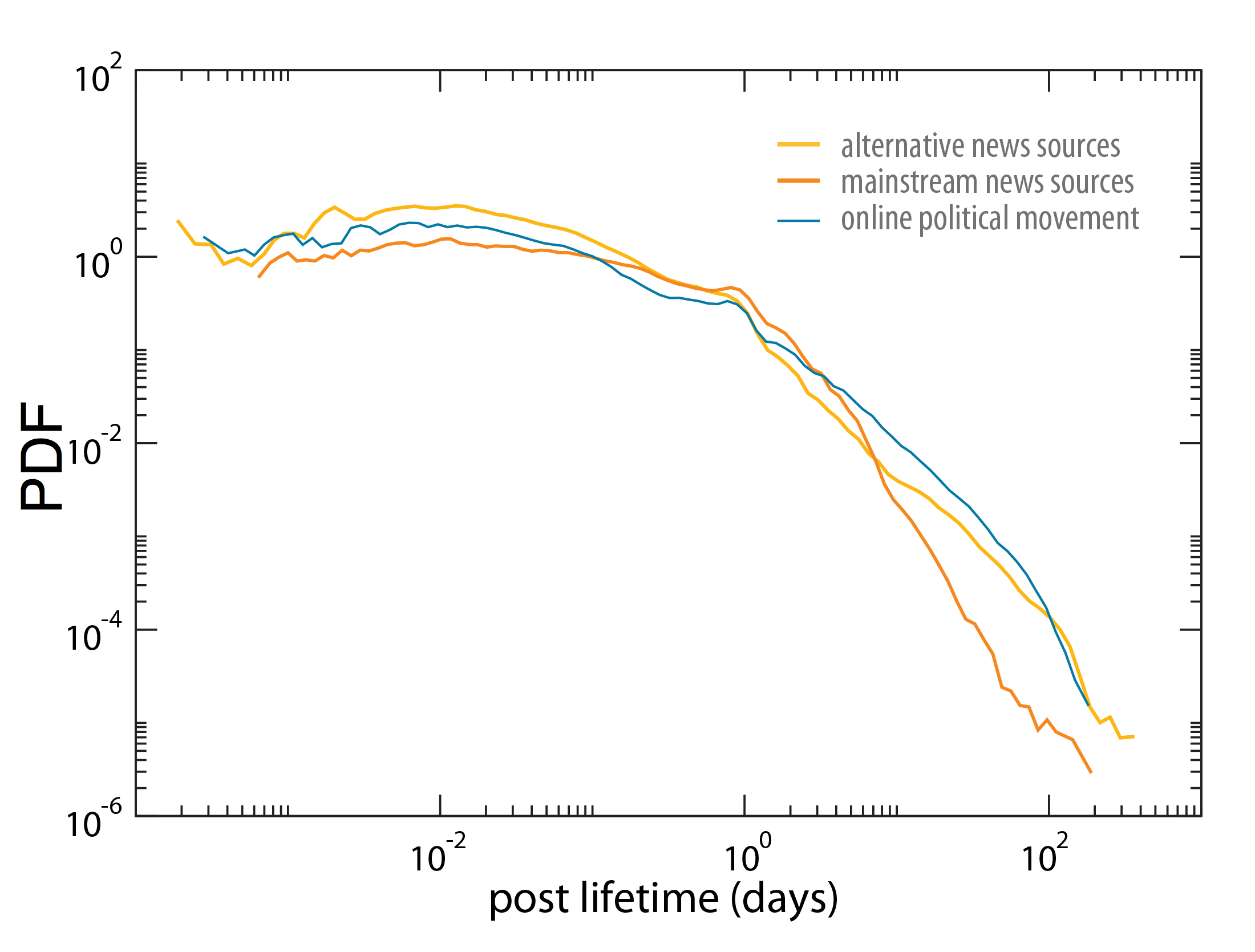} \label{fig:p_l}
\caption{\textbf{Post lifetime}. Probability density function, grouped by page type, of the temporal distance between the first and last comments of the post. Posts with qualitatively different topics (alternative information, political activism, and main stream news) show a similar behavior. }
\label{fig:lifetime_distribution}
\end{figure}

In Figure \ref{fig:lifetime_distribution} we show, for each page type, the probability density function of the post interest lifetime. This measure is computed as the temporal distance between the first and last comment of the post. 
Collective debates persist similarly (see Supporting Information section for further details), independently of whether the topic is the product of an official or unofficial source. 


Given the social context of these groups, what potential does a hurtful meme harness? In other words, how significant is the concurrent presence of users between different pages and how strong is the overlap?  Starting from the null-hypothesis, that each user has neither affiliation nor preference, we investigate the interaction dynamics of the different communities by quantifying the users that are present in both spaces.
The result in Table \ref{tab:common_users} hints that indeed a considerable number of users interact with pages of different classes.
The political discussion and alternative news get informed dominantly from each other rather than from mainstream media, while users of the first two sets are almost equally represented within the followers of mainstream newspapers.

\begin{table}[H]
\centering
    \begin{tabular}{l|l|c|clcl}
    \bf {Class A}  & \bf {Class B} & \bf {Common users (AB)} & \bf {Ratio (AB/A)} & \bf {Ratio (AB/B)} \\ \hline 
    Political movement & Alternative news  & $ 360,054 $ & $ 28.0\%$  & $ 33.6\%$ \\ \hline 
    Political movement & Mainstream news & $ 254,893 $ & $ 19.8\%$  & $ 32.4\%$ \\ \hline 
   Mainstream news & Alternative news & $ 278,337 $ & $35.4\%$  & $ 25.9\%$
	\end{tabular}
  \caption{ Common users between classes of pages. Many users make active contributions on pages with very different profiles. Several members of the political discussion are involved on both alternative news  pages and main stream newspapers pages in comparable ways.} 
  \label{tab:common_users}
\end{table}

In this portion of the Italian Facebook ecosystem untruthful rumors spread and trigger debate, representing an important part of the information flow animating the political scenario.

\subsection*{Response to false information}

Above results reveal that users consume unsubstantiated and main stream news in similar ways. In our study, both are consumed by users of political activism pages. Continuing our investigation, we want to understand if this information context might affect the users' selection criteria.
Therefore, we measure the reaction of users to a set of 2788 false information injected by a troll page - i.e, a page promoting caricatural version of alternative news and political activism stories.

In order to perform this analysis, we applied a classification strategy aimed at discriminating typical users for each one of the three categories of pages.
In particular, we were interested in distinguishing users based on their behavior. Having access to the 6 months historical likes, comments, and \textsl{likes to comments} on all posts within the timeframe (and within the privacy restrictions), we quantify the interaction of each user with the posts in each class. As we did this, the following assumptions were in place:

\begin{itemize}
\item The topic of the post is coherent with the theme of the page on which it was published.
\item A user is interested in the topic of the post if he/she \textit{ likes} the post. A comment, although it reflects interest, is more ambiguous, and, therefore, is not considered to express a positive preference of the topic. 
\item We neither have access to nor try to guess the page subscription list of the users, regardless of their privacy settings. Every step of the analysis involves only the active (participating) users on each page.
\end{itemize}

According to these assumptions, we use solely the \textit{likes} to the posts. For instance, if a user likes 10 different posts on one or multiple pages of the same political movement, but that user never liked posts of any other topic, we will \textit{label} that user to be associated with the political movement. 
Since it is not always the case that there is a clear preference, we have to take into account the random sampling bias error - since our data set represents indeed a sample of the users' Facebook activity. Given the limitations of the API, the only information we have about the user is how that user interacted with the posts we have downloaded.  

The labeling algorithm for each user is to calculate the 95\% confidence interval of percentage of likes of posts in each topic. Only if the confidence interval of the preferred topic does not overlap the other two topics, we assign the user a label. Although the true affiliation of the individual behind the end user can be a subjective matter, we believe that filtering out versatile users allows us to focus precisely on the rare, and more interesting, cases of interaction between highly polarized users.

\begin{figure}[H]
 \centering
       \includegraphics[width=.99\textwidth]{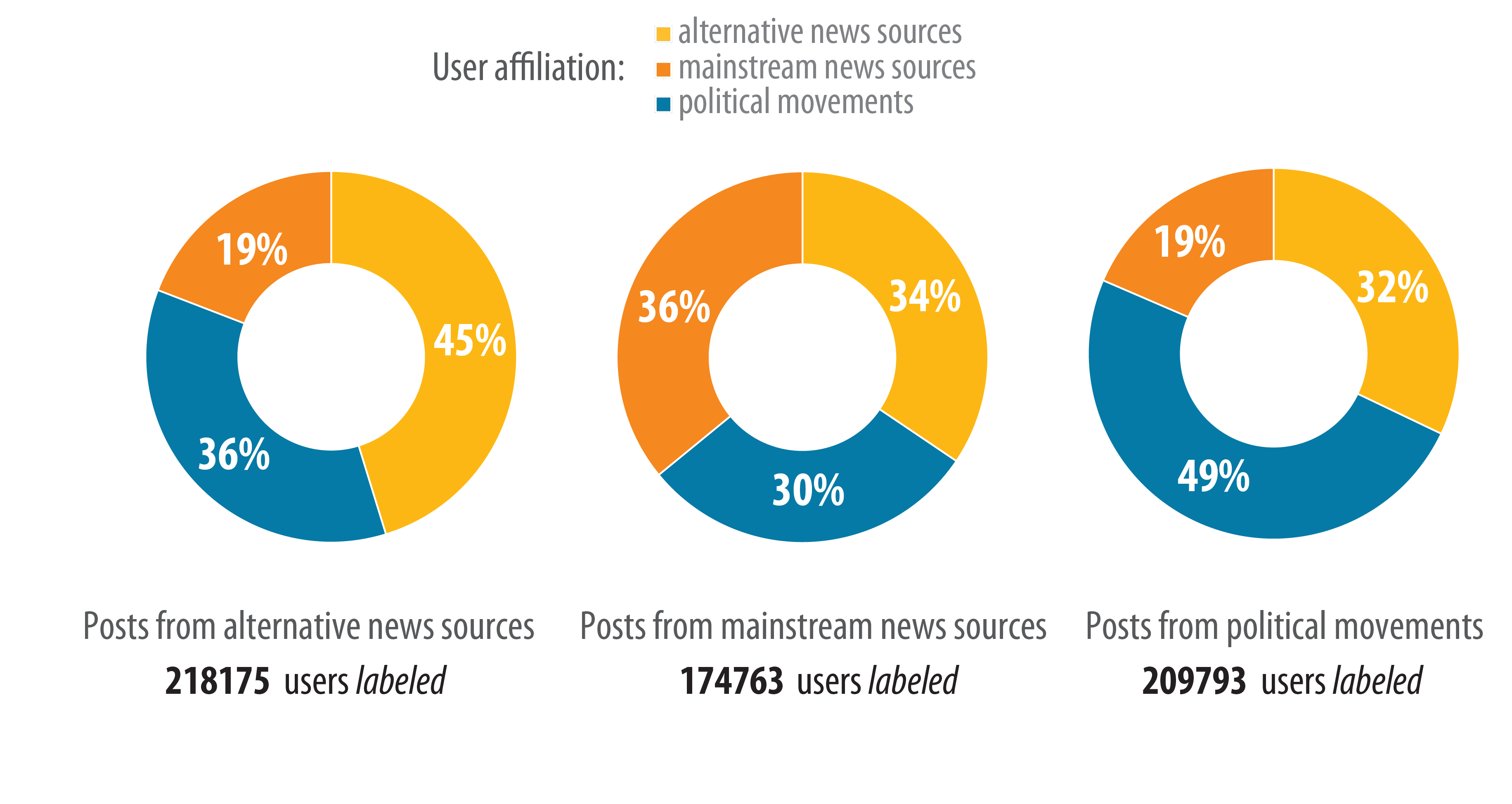} \label{fig:p_l}
\caption{For each page type: fractions of users with strong affiliations. }
\label{fig:pie_chart1}
\end{figure}

In Figure \ref{fig:pie_chart1} we illustrate for each page type, the respective contributions brought by labeled (polarized) users. It is important to note that this measure is not designed to describe the overall affiliation of the members of the page.
The fractions are computed by taking all the posts from a class and counting percentage of users coming from each profile.
Posts from alternative information sources and political activism pages present a clear supremacy of the predominant class of users with, respectively,  45\% and 49\% of the dominant class. 
Not surprisingly, mainstream media pages, present a more balanced distribution of user classes, as their purpose is to communicate neutral information.
However, users labeled as political activists are more active on alternative information pages than on mainstream newspapers. In turn, users labeled as mainstream media adepts are in minority on both alternative and activist pages. According to this partitioning of the information space, now we are able to distinguish interactions occurring between users pertaining to different regions of the ideological space.

Given the outline of users distribution within the various classes, we want to see which users are more responsive to the injection of false information in terms of interaction.
As before, we cannot use the comments as discriminators, as they can represent either positive or negative feedbacks with respect to the published topic. 
Therefore, we focus only on the users liking 2788 \textsl{troll} posts.

\begin{figure}[H]
 \centering
       \includegraphics[width=.99\textwidth]{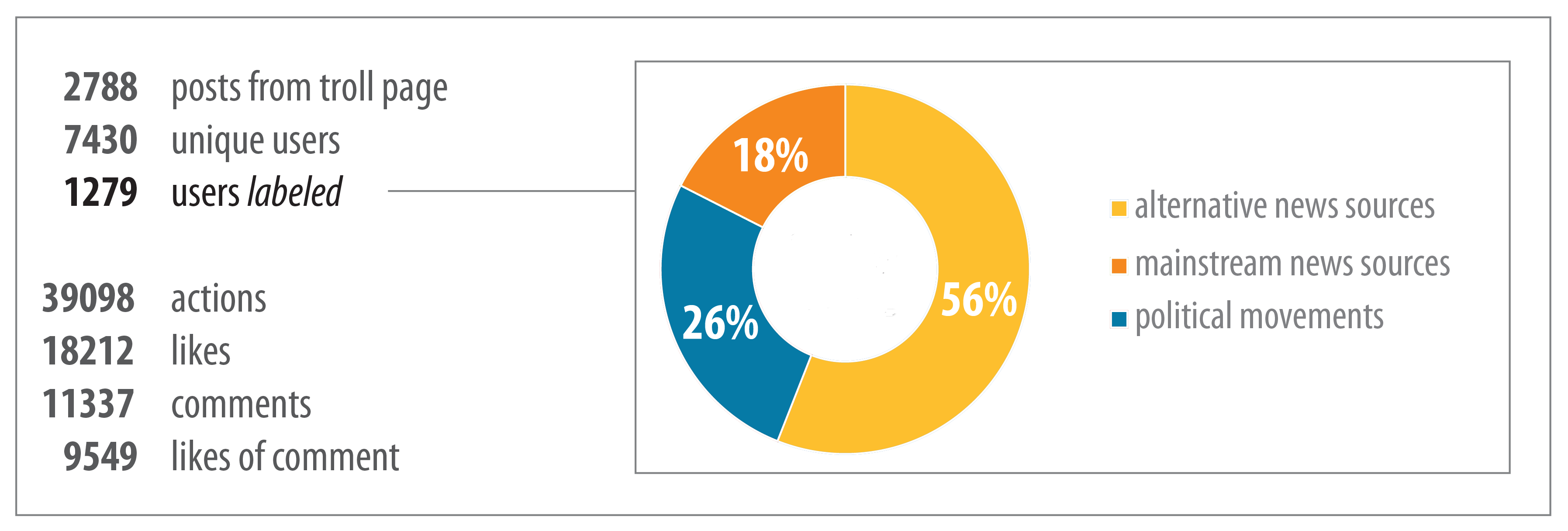} \label{fig:p_l}
\caption{Social response to intentionally injected false information. Labels represent user affiliation. The users more responsive to the injection of false information are the ones having strong affiliation alternative information sources.}
\label{fig:pie_chart2}
\end{figure}

As previously mentioned, \textsl{troll} posts are related to arguments debated by political activists or on alternative information sources but with a clear parodistic flavor.
For instance, one of the most popular memes that explicitly spread a false rumor (in text form) reads: \textit{Italian Senate voted and accepted (257 in favor and 165 abstentions) a law proposed by Senator Cirenga aimed at funding with 134 billion Euros the policy-makers to find a job in case of defeat in the political competition}.  We were able to easily verify that this meme contains at least four false statements: the name of the senator, the total number of votes is higher than possible, the amount of money (more than 10\% of Italian GDP) as well as the law itself. 
This meme was created by a troll page and, on the wave of public discontent against italian policy-makers, quickly became viral, obtaining about 35,000 shares in less than one month. Shortly thereafter, the image was downloaded and reposted (with the addition of a commentary) by a page describing itself as being focused on political debate. Nowadays, this meme is among the arguments used by protesters manifesting in several Italian cities. This is a striking example of the large scale effect of misinformation diffusion on the opinion formation process. 
As shown in Figure \ref{fig:pie_chart2} by counting the polarized users that liked the posts, we find that the most susceptible users to interact with false information are those that are mostly exposed and interacting with unsubstantiated claims (i.e. posts on alternative information pages). 

According to our results, users with strong preferences for alternative information sources, perhaps motivated by the will to avoid the manipulation played by mainstream media controlled by the government, are more susceptible to false information.

\section*{Conclusions}
Conspiracists generally tend to explain a significant social or political aspect as a secret plot by powerful individuals or organizations \cite{Sunstein12} and their activity is proliferating over the web. This study provides a genuine outline of the online social dynamics and, in particular, on the effect of Facebook on bursting the diffusion of false beliefs when truthful and untruthful rumors coexist.

In this work, we perform a case study aiming to understand the interlink between political discussion and information on the web. The portion of Facebook we analyzed presents a complex set of social interactions. 
Several cultures coexist, each one competing for the attention of users. Specifically, we observe a strong interaction between political discussion and information sources (either alternative or main stream).
Most of the online activism Facebook pages contain claims that mainstream media is manipulated by higher entities (and thus the information is be not neutral or reliable). 
Such an antagonism makes any kind of persuasion process, even if based on more solid information, very difficult.  
As a response to partisan debates, the emergent groups of \textsl{trolls} began to provide parodistic imitations of a wide range of online partisan topics.
Despite the evident parodistic (and sometimes paradoxical) contents, not rarely, troll memes fomented animated debates and diffused through the community as any other information would.
Through statistical analysis, we find that the consumption patterns are similar despite the different nature of the information. 
Finally, in order to uncover more characteristics of the process, we distinguished users with strong affiliations and observed their respective interaction patterns, as well as with false information inoculated in that portion of the Facebook ecosystem. 
We find that, out of the 1279 labeled users interacting with the troll memes, a dominant percentage (56\% , as opposed to $26\%$ and $18\%$ for other groups) is constituted of users preeminently interacting with alternative information sources and thus more exposed to unsubstantiated claims.
The results of our study raise a real warning, as the higher the number of circulating unsubstantiated claims is, the more users will be biased in selecting contents.


\section*{Acknowledgments}

Funding for this work was provided by the authors' institutions (IMT Lucca Institute for Advanced Studies, Northeastern University), EU FET project MULTIPLEX nr.317532. The funders had no role in study design, data collection and analysis, decision to publish, or preparation of the manuscript.

We want to thank Alessandro Vespignani, Rosaria Conte, Mario Paolucci, Santo Fortunato, Brian Keegan, Piotr Sapiezynski and Gianni Riotta for useful discussions 

Special thanks go to Giulia Borrione, Dino Ballerini, Elio Gabalo, Monica D'Agruma, Stephanie Ponzo, Giacomo Sorbi, Alpi Stefano, Umberto Mentana, Salvatore Previti for pointing out the phenomenon of misinformation and for precious suggestions.

\section*{Authors Contribution}

Conceived and designed the experiments: WQ DM. Performed the experiments: WQ. Analyzed the data: WQ DM QZ. Contributed reagents/materials/analysis tools: DM LR QZ. Wrote the paper: WQ DM LR QZ MK.

\bibliography{trolling}

%
%

\newpage
\section{Supporting Information}

\subsection*{Data Collection}

The entire data collection process is performed exclusively with the Facebook Graph API, which is publicly available and which can be used through one's personal Facebook user account.

The pages from which we download data are public Facebook entities (can be accessed by virtually anyone). Most of the user content contributing to such pages is also public unless the user's privacy settings specify otherwise.

Using the Graph API, one is able to download the \textsl{feed} of a page, along with the content that would otherwise be visible though regular browsing. There are, however, several limitations to this process which we will attempt to describe in detail later in this section.

Our approach: for each page we access the feed between Sept 1st, 2012 and Feb 28th, 2013; that is, we download all the public posts created in this period and the related content. Specifically, we can identify the users that liked the post, we can access the comments (content, user, time). Other fields such as post author, post creation time, and post type (e.g. photo, link, video, plain text) are also available. 

Note 1: When a post points to a photo, it has two possible forms. The first is when the photo upload is the post itself. The second type is when the post actually links to a photo that already existed as a static object (i.e. it pertains to an album and has a fixed \textsl{address} as well as a unique identifier). In the latter case, if one were to click on the post, one is redirected to the unique address of the photo in case. Therefore, it is possible to have different posts pointing to the same fixed object (photo or video). It is important at this point to make the distinction between post and object. A post can accrue comments, likes, or shares that are separate from the comments, likes, and shares of the object it points to. For this reason, when we encounter a post that also points to an object, we download the data associated with the latter as well. Naturally, an object has the benefit of having accrued more interest over time.

The rate limits that we encounter with the Graph API restrict us to only being able to access the last 5000 shares of an object or post. The likes and comments, however, are not subject to such rate limits.

At all times the privacy limits are in effect, so the data we obtain is usually smaller than the total number theoretically available. For instance, it is possible that we only get  about 4000 of the last shares of an object. Similarly, we can only get partial data on likes and comments. Since we do have access  to the total number of shares, likes, comments, according to our observation, about 20\% of the user actions are invisible to us. This property is specific to our users/set of pages and does not necessarily accurately reflect the properties of the rest of the Facebook community outside our dataset.
  
Note 2: Within the privacy restrictions, one is able to access the branches of each share action, should they exist. The reverse process (upstream) is not possible using the Graph API, unlike the case of direct web surfing. For this reason, along with the privacy settings of some users, it is virtually impossible to reconstruct the complete sharing tree of a popular object (with more than 5000 shares). 

Note 3: One important limitation of the Graph API, which manual surfing does not present, is that one cannot access any user profile information, even if such information is otherwise public. Consequently, we only have access to the unique id and to the name of each user - but not to their location, for instance.

\subsection*{Attention patterns: likes and comments distributions }

\subsection*{Post lifetime in the different classes}

\begin{figure}[H]
 \centering
       \includegraphics[width=.6\textwidth]{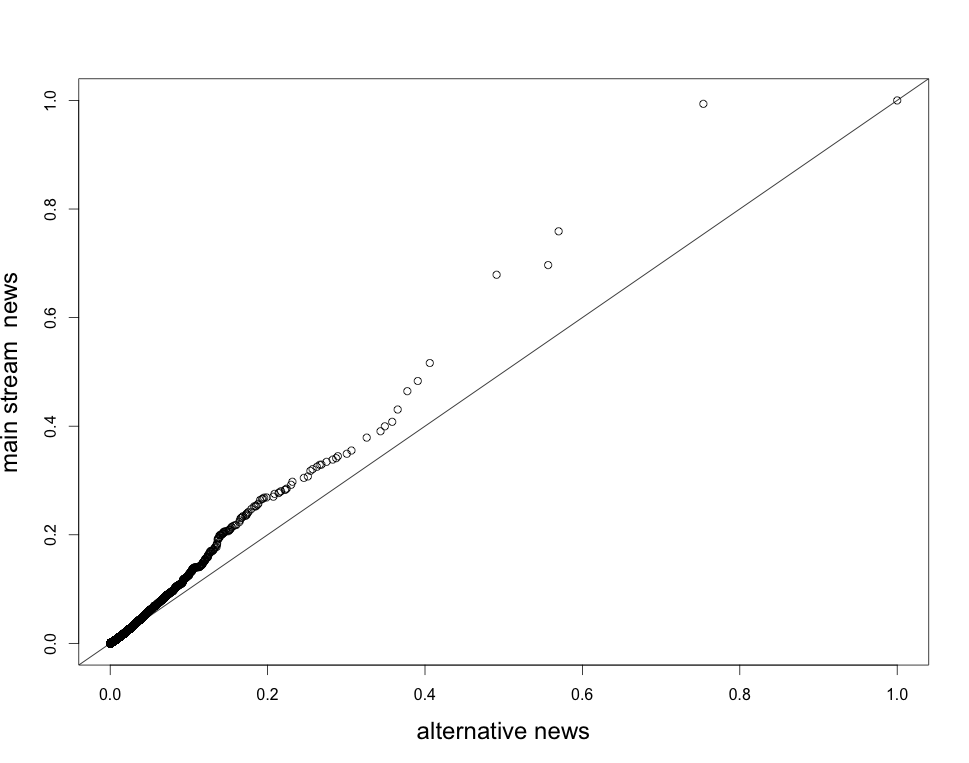} \label{fig:p_l}
\caption{Quantile vs Quantlie of post lifetime in alternative news and mainstream news.}
\label{fig:qq_alt_control}
\end{figure}

\begin{figure}[H]
 \centering
       \includegraphics[width=.6\textwidth]{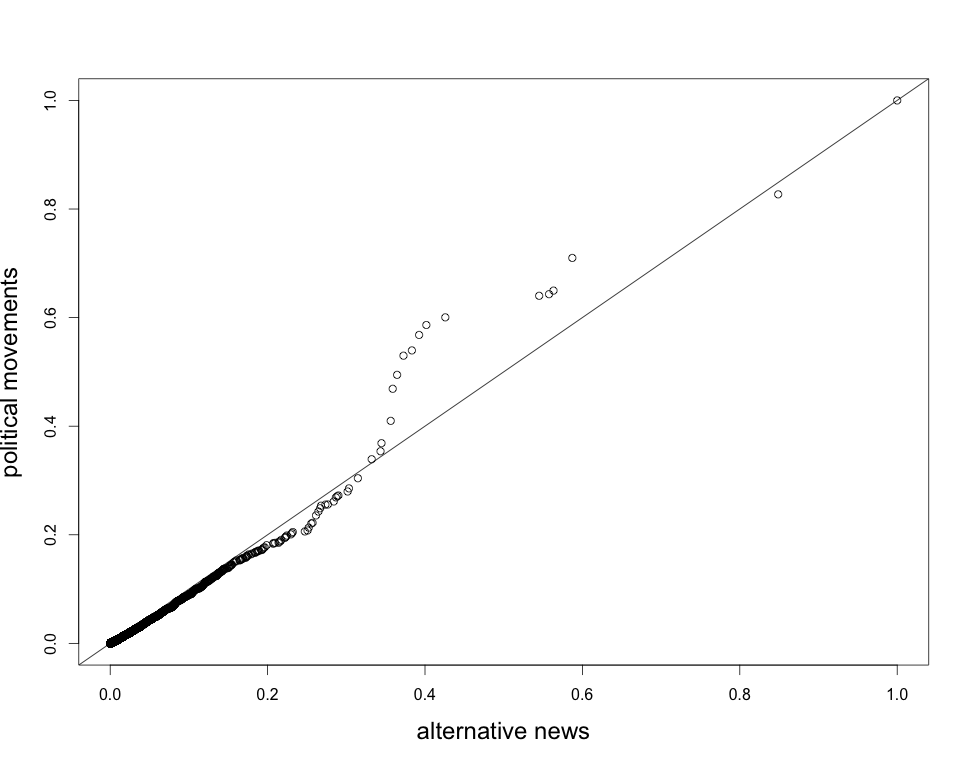} \label{fig:p_l}
\caption{Quantile vs Quantlie of post lifetime in alternative news and political movements.}
\label{fig:qq_alt_pop}
\end{figure}

\begin{figure}[H]
 \centering
       \includegraphics[width=.6\textwidth]{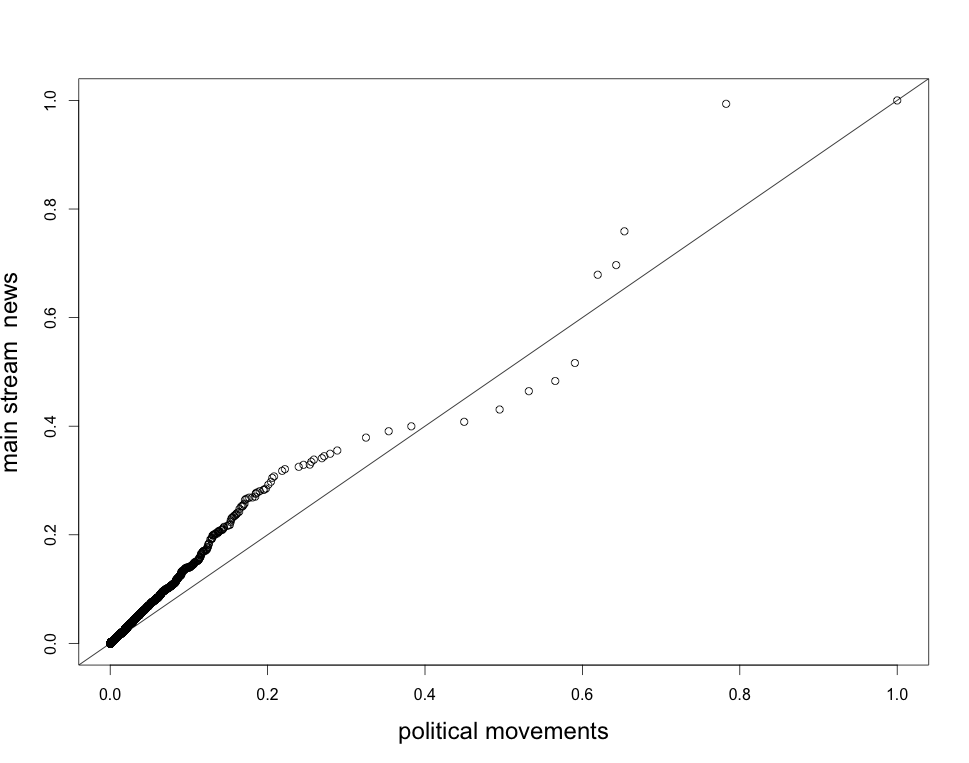} \label{fig:p_l}
\caption{Quantile vs Quantlie of post lifetime in mainstream news and political movements.}
\label{fig:qq_control_pop}
\end{figure}

In this section we provide further details about the post lifetime in the various classes.
We consider lifetime as the distance in time between the first and the last comment to a post.
Here we provide a comparisons among the distributions of the lifetime for each post according to the various classes.
In Figures \ref{fig:qq_alt_control}, \ref{fig:qq_alt_pop} and \ref{fig:qq_control_pop} we show the quantile-quantile plot of the normalized distributions.

In Table \ref{tab:ks} we report the results of Kolmogorov-Smirnov test.
Notice that since the KS test loses precision in heavy tailed data we perform the analysis by cutting the tail at  $0.1$ on the normalized distributions.

\begin{table}[H]
\centering
    \begin{tabular}{l|l|c|}
    \bf {Class A}  & \bf {Class B} & \bf {p-value}\\ \hline 
    Political movement & Alternative news  & $ 0.9253 $  \\ \hline 
    Political movement & Mainstream news & $ 0.1087 $ \\ \hline 
   Mainstream news & Alternative news & $ 0.156$
	\end{tabular}
  \caption{Kolmogorov-Smirnov test on the different distributions coming from the different classes.} 
  \label{tab:ks}
\end{table}

In Table \ref{tab:ks} we report the results of t test.
As in the KS case we cut the tail at $0.1$ of the normalized distributions.

\begin{table}[H]
\centering
    \begin{tabular}{l|l|c|}
    \bf {Class A}  & \bf {Class B} & \bf {p-value}\\ \hline 
    Political movement & Alternative news  & $ 0.3249 $  \\ \hline 
    Political movement & Mainstream news & $0.9863 $ \\ \hline 
   Mainstream news & Alternative news & $ 0.2768 $
	\end{tabular}
  \caption{T-test on the different distributions coming from the different classes.} 
  \label{tab:ks}
\end{table}

\newpage
\subsection*{List of pages}



\begin{tabularx}{\textwidth}{ |p{3cm}|p{4cm}| p{1.5cm}|X| }
\hline 
Category & Page & Followers & Description\tabularnewline
\hline 
\hline 
Alt.Inf & Libertà di Stampa & 289k & Supporting the right of expression. Reporting news mostly oriented
on politics (lobbies and government). \tabularnewline
\hline 
Alt. Inf & Lo Sai Economia & 7k & Delivering news about economics (lobbies, signorage, free masons and
so forth)\tabularnewline
\hline 
Alt. Inf & Informazione Libera & 1m & Pointing out all the corruption of the political ``lobby'' and the
current dramatic situation of the middle class\tabularnewline
\hline 
Alt. Inf & Il Radar & 15k & News Magazine supporting the Right party. Pointing out all the paradox
(sometimes exagerated) of the opposite parties \tabularnewline
\hline 
Alt. Inf & NeoVitruvian & 5k & NWO order, the miracle of the free energy by Nicola Tesla, Illuminati
etc etc\tabularnewline
\hline 
Alt. Inf & NoCensura.com & 480k & All the things that are ``intentionally ignored by manipulated media''
(signorage, opinion manipulation..)\tabularnewline
\hline 
Alt. Inf & Lo Sai Salute e Alimentazione & 18k & Information and discussion against the traditional medicine practices
(Vaccines and Autism, OGM, etc)\tabularnewline
\hline 
Alt. Inf & Terra Real Time & 38k & Signorage, Alternative healtcare, David Icke thesis, Chemical Trails...\tabularnewline
\hline 
Alt. Inf & Lo Sai Chemtrails & 6k & The cronichle of the chem trails....\tabularnewline
\hline 
Alt. Inf & Signoraggio Bancario & 6k & All news and info about the advances of the NWO plans and the world
dominated by the banks (main thesis BCE and Federal Reserve are private
companies creating the public debt with an escamotage)\tabularnewline
\hline 
Alt. Inf & Nuovo Ordine Mondiale & 3k & Discussion about the NWO plans and its evidences\tabularnewline
\hline 
Alt.Inf & L'altra Notizia & 70k & All the information neglected by the manipulated media\tabularnewline
\hline 
Alt. Inf & Contro Copertina & 10k & Mainly populistic arguments to convey public discontent by the wave
of indignation\tabularnewline
\hline 
Alt. Inf & Haarp Controllo Globale & 3k & Diffusion of information about the existence of secret plans to destabilize
the world by causing earthquakes and by poisoning humanity with barium
and other not well specified substancies released by airplanes. \tabularnewline
\hline 
Alt. Inf & Stop alle Scie Chimiche & 14k & Pointing out the Chemical Trails plans\tabularnewline
\hline 
Alt. Inf & LoSai.Eu & 135k & One of the most active pages in disseminating all information neglected
by main stream media\tabularnewline
\hline 
Alt.Inf & Verità 11 Settembre & 14k & Supporting alternative thesis about 9/11 official version\tabularnewline
\hline 
Alt.Inf & Uniti contro le multinazionali & 17k & Major corporation are poisoning the world and natural medications
has to replace medicines (curing cancer with bicarbonatum)\tabularnewline
\hline 
Alt.Inf & Condividi la Conoscenza & 95k & News to about all the facts (mainly politics) neglect by main stream
media\tabularnewline
\hline 
Alt. Inf & Informare per resistere & 762k & One of the most active page to diffuse information neglected by the
media through the web\tabularnewline
\hline 
Alt. Inf & Contro Informazione Alternativa & 45k & News with a populistic tone in particular against the political actions
of the government\tabularnewline
\hline 
Alt.Inf & Orwell 2012 & 2k & NWO plans and opinion manipulation\tabularnewline
\hline 
Alt.Inf & HAARP & 2k & Diffusion of information about the existence of secret plans to destabilize
the world by causing earthquakes and by poisoning humanity with barium
and other not well specified substancies released by airplanes. \tabularnewline
\hline 
Alt. Inf & Contro l'informazione manipolata & 103k & All the information neglected by the manipulated media controlled
by the governments and lobbies\tabularnewline
\hline 
Alt. Inf & Informare Contro Informando & 32k & News to about all the facts (mainly politics) neglect by main stream
media\tabularnewline
\hline 
Alt.Inf. & Smascheriamo gli illuminati & 3k & Pointing out all the symbols and subliminal messages from the main
stream media delivered by the illuminati\tabularnewline
\hline 
Troll & Semplicemente me & 5k & Diffusing memes as a paradoxal imitation of alternative information
and political movements\tabularnewline
\hline 
Main Stream Media & Il Giornale & 125k & National official news paper near a the center right party (PDL) of
Berlusconi\tabularnewline
\hline 
Main Stream Media & La Repubblica & 1.3m & National news paper close to the center left party (PD) and the most
diffused italian journal\tabularnewline
\hline 
Main Stream Media & Il Fatto Quotidiano & 1.1m & National news paper near at the Five Star Movement (M5S) of Beppe
Grillo\tabularnewline
\hline 
Main Stream Media & Il Manifesto & 69k & National news paper near to the left party\tabularnewline
\hline 
Main Stream Media & Il Corriere della Sera & 1m & National News Paper\tabularnewline
\hline 
Main Stream Media & La Stampa & 131k & National News Paper\tabularnewline
\hline 
Main Stream Media & Il Sole 24 ore & 267k & National News Paper more oriented on the economic and financial systems\tabularnewline
\hline 
Main Stream Media & Il Messaggero & 183k & National News Paper\tabularnewline
\hline 
Political Activism & Incazzati contro la casta & 78k & Convey the public discontent against the socio-economic situation\tabularnewline
\hline 
Political Activism & RNA-Rete Anti Nucleare & 47k & Inform and sensibilization against the Nuclear Energy in Italy\tabularnewline
\hline 
Political Activism & Indignados Italia & 44k & Discussion about the current socio-economic situation\tabularnewline
\hline 
Political Activism & Questa è l'Italia & 118k & Convey the public discontent against the socio-economic situation\tabularnewline
\hline 
Political Activism & NewApocalypse & Not Online & Convey the debate against Signorage and the NWO \tabularnewline
\hline 
Political Activism & Qelsi & 90k & Against the left parties \tabularnewline
\hline 
Political Activism & Partigiani del III Millennio & Not more online & Convey the public discontent against the socio-economic situation\tabularnewline
\hline 
Political Activism & Vogliamo i Parlamentari in Carcere... & 45k & Convey the public discontent against the socio-economic situation
(arresting all policy makers)\tabularnewline
\hline 
Political Activism & Adesso Fuori dai Coglioni & 560k & Convey the public discontent against the socio-economic situation\tabularnewline
\hline 
Political Activism & Forza Nuova & 50k & Convey the public discontent against the socio-economic situation
(extreme right party)\tabularnewline
\hline 
Political Activism & Vota Casa Pound & 48k & Convey the public discontent against the socio-economic situation
(extreme right party)\tabularnewline
\hline 
Political Activism & Gruppo Free Italy & 15k & Convey the public discontent against the socio-economic situation\tabularnewline
\hline 
Political Activism & Catena Umana Attorno al Parlamento  & 105k & Convey the public discontent against the socio-economic situation\tabularnewline
\hline 
Political Activism & Non chiamateli Politici ma criminali & 45k & Convey the public discontent against the socio-economic situation\tabularnewline
\hline 
Political Activism & BeppeGrillo & 1m & The page of the M5S Leader proposing the free circualtion of the information
on the Internet as a major revolution. A good example of e-participation.\tabularnewline
\hline 
Political Activism & Alice nel paese delle merdaviglie & 70k & Convey the public discontent against the socio-economic situation\tabularnewline
\hline 
\end{tabularx}

\end{document}